# Energy Harvesting in M2M and WSN Space

Sugata Sanyal

Corporate Technology Office, Tata Consultancy Services, India

sanyals@gmail.com

**Abstract:** Energy harvesting or power harvesting or energy scavenging is a process where energy is derived from external sources (e.g. solar power, thermal energy, wind energy, salinity gradients, kinetic energy etc.), captured, and stored for small, wireless autonomous devices, like those used in wearable electronics and wireless sensor networks. This paper is focused to applications of Energy Harvesting in Wireless Sensor Networks. This is going to help the ever growing M2M (Machine to Machine) field where there is an exponential growth of intelligent devices and automatic control of these is of paramount importance. [1]

**Energy Harvesting: A Survey:** Wireless Sensor Nodes (WSN) is used for various purposes. One of the biggest disadvantages of WSN is that these are mostly used in remote places and in unattended fashion. These at present are powered by small batteries and due to continuous usage batteries drain within some time. Replacing these batteries is mostly difficult if not impossible. Scientists have been looking for some solution to this issue and WSN with capacity of working on harvested energy from various resources evolved out of this requirement.

Coarasa et al. [2] has proven that mobile energy sources, which are called *actors*, could be moved along a pre-decided pattern, which can radiate appropriate level of electro-magnetic energy which will be sufficient to charge the WSNs. The work has calculated all aspects of energy transfer between the *actor* and the *sensor*. This result has also been verified by extensive simulation.

Schmidt [3], in his article in Wireless Communications, has provided a cohesive set of details about Energy Harvesting Details and its relevance to Machine to Machine Communications (M2M).

It has been predicted that Internet connected devices will increase from an approximate 300 Million (by 2017) to 50 Billion (by 2020) in number. The issue of controlling such a huge number of intelligent devices is a daunting task. And in most places, it is assumed that M2M communication will be automatic to make it a viable proposition. We observe that development and enhancement of cloud services will help in enhancing the M2M operations.

To empower this huge number of devices, power support provided by standard battery will become increasingly difficult. Moreover, many times, the WSN devices are strewn across remote fields where the result from these active devices is useful, but once they die, accessing and battery replacement is almost impossible. Energy harvesting will provide continuous power to the new generation of battery-less WSNs. These will provide continuous service from these WSNs and it will also need no maintenance. For critical applications, it is possible to use a technique called COUNCIL based clusters of WSNs [4]. Here the technique used is as follows. Simply put, as a Single Cluster is apt to fail more, a group of Clusters, called COUNCIL, is used to replace a single cluster. Service of the single cluster head is replaced by an Average of multiple cluster heads which uses a threshold secret sharing method. This provides better security. Again, all elements (WSN) are powered by energy transfer through harvesting.

Energy may be harvested by either solar energy or by miniaturized electrodynamic energy converter. A thermoelectric converter can produce energy from heat. We may add a capacitor to store energy when ambient energy is not available. Sub-GHz signal is tapped for harvesting energy from RF sources and it is also to be noted that this provides a higher range of 300 m outside and 30 m inside a building.  For Building Automation, energy harvesting WSNs are heavily used, thus eliminating costly battery replacement. Here, controlling of light, metering, energy management etc. are the main targeted area for automation. In this application, all control elements get information about various parameters from various battery-less sensors and these are passed on to the central Building Automation System.

Self-powered radiator valves are automatically controlled by energy which is generated from the temperature difference between surrounding air (in ambient temperature) and hot water which in turn controls the connectivity with a controller and also for controlling the valve. These systems thus have no cable connection, no batteries and needs zero maintenance. HVAC (Heating, Ventilation and Air-Conditioning) control is a flexible automation system where a Thermostat, VAV (Variable Air Volume) or Fan Coil Controller are the recipient of information related to occupancy, temperature, humidity, window position or $CO_2$ from the corresponding battery-less sensors and this in turn controls the valve actuators for radiators or dampers for the VAV systems. This controller is also connected to a Building Automation System thus providing a centralized control of the building parameters over Internet.

Alarm systems e.g. water detector or fire detector will have more stringent requirement for its continuous flawless working. Battery-less sensors are of great help here as these systems are inherently more reliable as they do not have any battery (which will fail) or wiring system. Energy harvesting sensors play a big role here.

All system parameters could be passed through some communication protocol (in encrypted format) to a remote Cloud environment.  All related management systems could gather system parameters through this cloud-based system and could pass on control information through this system, as well.  When cloud-based systems are used for sensitive control systems, it is necessary to take care of the cloud security issues [5], [6].

**Future Aspects of Energy Harvesting**:  Advanced Ultra-Low power Wireless Communication has progressed to a great level and this in turn has helped in advanced M2M applications.  Smart Home, Smart Metering and Smart Grid are few areas where M2M is already playing a big role through application of Energy Harvesting. Other areas like logistics for Industry and Transportation are having advanced research on the anvil, where M2M using energy harvesting is active.

**Some unusual applications:**  Though not directly connected with WSN and M2M technology, research is ongoing for bio-medical devices (pace-maker, middle-ear implant etc.) where the device may harvest energy through the body movement itself [7]. These are in deep research domain. One example:  A person with Pace-Maker with an Energy Harvesting Technology and no battery will last for many years. In case, this person gets a heart attack, his movement will stop, so no power to his pacemaker. So a hybrid technology will need to be evolved where such pace-makers will also have a battery backup. In case of accidental stoppage of movement, this battery will start providing power to the pace-maker. Normally, this battery will keep getting charged for many years.

References:
[1]   Energy Harvesting (from Wikipedia) at http://en.wikipedia.org/wiki/Energy_harvesting

[2]  Impact of Mobile Transmitter Sources on Radio Frequency Wireless Energy Harvesting: A. H. Coarasa, P. Nintanavongsa, Sugata Sanyal, Kaushik R. Chowdhury; 2013 International Conference on Computing, Networking and Communication - Green Computing, Networking and Communications Symposium, January, 2013, San Diego, USA, pp. 1-5, IEEE Com, Draft Copy: http://arxiv.org/abs/1312.0883

[3] Energy harvesting wireless: the secret to M2M's success:  April 08, 2013 | Frank Schmidt, EnOcean | 222903833,
Wireless Communications at http://www.microwave-eetimes.com/en/energy-harvesting-wireless-the-secret-to-m2m-s-success.html?cmp_id=71&news_id=222903833

[4] Forming the COUNCIL based clusters in securing WSNs by A. Ojha, H. Deng, S. Sanyal, D. P. Agrawal, 2004, International Conference on Computers and Device Communication (CODEC 2004), Draft Copy: http://arxiv.org/abs/1003.2614

[5] Survey on Security Issues in Cloud Computing and Associated Mitigation Techniques by  Rohit Bhadauria, Sugata Sanyal, Draft Copy:  http://arxiv.org/abs/1204.0764

 [6] A survey on security issues in cloud computing by Rohit Bhadauria, Rituparna Chaki, Nabendu Chaki, Sugata Sanyal, http://arxiv.org/abs/1109.5388v1

[7] The Body Electric: How much energy can you extract from a dance? By Emma Roller
http://www.slate.com/articles/health_and_science/alternative_energy/2013/03/kinetic_energy_harvesting_technology_to_power_lights_cell_phones_medical.single.html

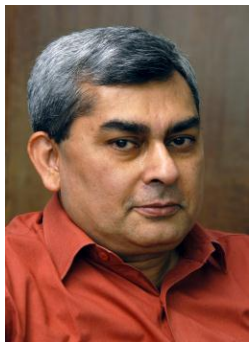

Dr. Sugata Sanyal was a Professor in the School of Technology & Computer Science at the Tata Institute of Fundamental Research. He is a Research Advisor with the Tata Consultancy Services now. Please visit the website for more.